\documentclass[final,5p,times,twocolumn,authoryear]{elsarticle}

\usepackage{amssymb}

\usepackage[breaklinks]{hyperref}
\usepackage{amsmath}
\usepackage{aas_macros}

\journal{Astronomy and Computing}

\graphicspath{{./},{./figures/}}
\newcommand{\degr}{\ensuremath{^\circ}}
\newcommand{\arcmin}{\ensuremath{^\prime}}
\newcommand{\arcsec}{\ensuremath{^{\prime\prime}}}

\begin{document}

\begin{frontmatter}



\title{\texttt{FLaapLUC}: a pipeline for the generation of prompt alerts on transient \textit{Fermi}-LAT $\gamma$-ray sources}


\author[jpl]{J.-P.~Lenain\corref{cor}\fnref{fn}}
\ead{jlenain@in2p3.fr}

\cortext[cor]{Corresponding author}
\fntext[fn]{ORCID: \href{http://orcid.org/0000-0001-7284-9220}{0000-0001-7284-9220}}

\address[jpl]{Sorbonne Universit\'es, UPMC Universit\'e Paris 06, Universit\'e Paris Diderot, Sorbonne Paris Cit\'e, CNRS, Laboratoire de Physique Nucl\'eaire et de Hautes Energies (LPNHE), 4 place Jussieu, F-75252, Paris Cedex 5, France}

\begin{abstract}
The large majority of high energy sources detected with \textit{Fermi}-LAT are blazars, which are known to be very variable sources. High cadence long-term monitoring simultaneously at different wavelengths being prohibitive, the study of their transient activities can help shedding light on our understanding of these objects. The early detection of such potentially fast transient events is the key for triggering follow-up observations at other wavelengths. A \texttt{Python} tool, \texttt{FLaapLUC}, built on top of the \textit{Science Tools} provided by the \textit{Fermi} Science Support Center and the \textit{Fermi}-LAT collaboration, has been developed using a simple aperture photometry approach. This tool can effectively detect relative flux variations in a set of predefined sources and alert potential users. Such alerts can then be used to trigger target of opportunity observations with other facilities. It is shown that \texttt{FLaapLUC} is an efficient tool to reveal transient events in \textit{Fermi}-LAT data, providing quick results which can be used to promptly organise follow-up observations. Results from this simple aperture photometry method are also compared to full likelihood analyses. The \texttt{FLaapLUC} package is made available on GitHub and is open to contributions by the community.
\end{abstract}


\end{frontmatter}

%

\section{Introduction}

The sky is not as immutable and quiet as it first seems when seen with the naked eye. Once studied in detail with sensitive instruments, variable sources are detected at all wavelengths and on different time scales. This notably applies at high energies, and more particularly to non-thermal emission from sources such as active galactic nuclei (AGN), pulsars, binaries, micro-quasars or cataclysmic variables.

The \textit{Fermi}-LAT $\gamma$-ray instrument \citep{2009ApJ...697.1071A} has revealed many new high energy sources \citep{2010ApJS..188..405A,2012ApJS..199...31N,2015ApJS..218...23A}, many of which are constituted by AGN \citep{2010ApJ...715..429A,2011ApJ...743..171A,2015ApJ...810...14A}. AGN are highly variable by nature, and a quick identification of any ongoing unusual activity in these sources is crucial to ensure multi-wavelength follow-up observations, to better characterize the nature of their emission.

A full picture of the behaviour of AGN could in principle be obtained with simultaneous, high-cadence monitoring at all available wavelengths. However, such observational campaigns are hardly practically achievable on long time scales. Instead, a grasp of knowledge can be picked up during flaring events, if several facilities follow up on the flare simultaneously. \textit{Fermi}-LAT is particularly useful to monitor the whole sky in the high energy range (from 20\,MeV up to 300\,GeV and above), with full sky snapshots obtained every three hours. In the case of transient events, a prompt reaction to trigger multiwavelength observations is essential.

The \textit{Fermi}-LAT collaboration developed the FAVA (Fermi All-sky Variability Analysis) tool \citep{2013ApJ...771...57A}, which has the huge advantage of blindly searching for transient events all across the sky, but the latency time of about one week before releasing the data\footnote{\href{http://fermi.gsfc.nasa.gov/ssc/data/access/lat/FAVA}{fermi.gsfc.nasa.gov/ssc/data/access/lat/FAVA}} prevents using it for prompt, quick alerts and subsequent observations before a flare subsides. The typical duration of flares in blazars is indeed shorter than a week \citep[see e.g.][]{2013A+A...559A.136H,2013A+A...554A.107H,2016ApJ...816...53W,2016ApJ...824L..20A,2016ApJ...830..162B,2017MNRAS.464..418R,2017ApJ...836..205A}. The \textit{Fermi}-LAT collaboration also provides to the community a set of light curves, updated daily, on a list of bright sources\footnote{\href{https://fermi.gsfc.nasa.gov/ssc/data/access/lat/msl_lc}{fermi.gsfc.nasa.gov/ssc/data/access/lat/msl\_lc}}, and the \textit{Fermi} Science Support Center makes available aperture photometry light curves\footnote{\href{https://fermi.gsfc.nasa.gov/ssc/data/access/lat/4yr_catalog/ap_lcs.php}{fermi.gsfc.nasa.gov/ssc/data/access/lat/4yr\_catalog/ap\_lcs} and \href{https://fermi.gsfc.nasa.gov/ssc/data/analysis/user/aperture.pl}{https://fermi.gsfc.nasa.gov/ssc/data/analysis/user/aperture.pl}} for all sources belonging to the 3FGL catalogue \citep{2015ApJS..218...23A}. However, for sources not reported in the 3FGL catalogue and/or if one wants to consider light curves with a different time binning, a custom pipeline is necessary.

Following major interests by the H.E.S.S. collaboration in target of opportunity (ToO) observations, \texttt{FLaapLUC} (\textit{\underline{F}ermi}-\underline{L}AT \underline{a}utomatic \underline{a}perture \underline{p}hotometry \underline{L}ight \underline{C}$\leftrightarrow$\underline{U}rve) has thus been developed, built in \texttt{Python} on top of the \textit{Science Tools} provided by the \textit{Fermi} Science Support Center and the \textit{Fermi}-LAT collaboration, and based on the simple aperture photometry technique \citep{2017ascl.soft09011L}. \texttt{FLaapLUC} has been extensively tried and tested within the H.E.S.S. collaboration. This tool is able to quickly analyse a pre-defined list of sources and automatically send alerts within H.E.S.S. in the case of sufficiently bright events occurring at high energies. This allows the H.E.S.S. collaboration to promptly trigger follow-up ToO observations. Moreover, an advantage of developing a custom tool such as \texttt{FLaapLUC} resides in the fact that the trigger criteria are under full control.

In the following, the method used for the \textit{Fermi}-LAT data analysis with \texttt{FLaapLUC} is described in Sect.~\ref{sec:method}. The performance, such as its false alarm trigger rate, and comparisons with classic full likelihood results, is discussed in Sect.~\ref{sec:perf}, before concluding in Sect~\ref{sec:ccl}.

\section{Description of the method}
\label{sec:method}

\texttt{FLaapLUC} uses the aperture photometry approach\footnote{\href{https://fermi.gsfc.nasa.gov/ssc/data/analysis/scitools/aperture_photometry.html}{fermi.gsfc.nasa.gov/ssc/data/analysis/scitools/aperture\_photometry}} to analyse \textit{Fermi}-LAT data. The aperture photometry is a simple method consisting in summing up photon counts in a region of interest, and weighting the results by the instrumental exposure to evaluate a flux. No background modelling or subtraction is performed. The goal of \texttt{FLaapLUC} is to provide alerts on ongoing activities in the \textit{Fermi}-LAT data from a predefined list of sources being monitored. Indeed, contrary to FAVA, to keep computing resources at a reasonable usage, a blind search of transient events across the full sky using aperture photometry is not performed.

The implementation of \texttt{FLaapLUC} is based on the standard \textit{Science Tools}. \texttt{gtselect} is used to extract the events around a source of interest, within a radius of 1\degr\ in the case of aperture photometry. This is because this method assumes that the data set is background-free. This assumption is of course wrong in the case of \textit{Fermi}-LAT data which are highly contaminated by Galactic and extragalactic diffuse emission. However, these diffuse components are not supposed to vary, and their presence will not impede detecting relative flux variations. The considered energy range is 100\,MeV--500\,GeV, and a cut on the maximal zenith angle of 90\degr\ is applied, as recommended by the \textit{Fermi}-LAT collaboration for point-source (event class 128) analyses using the Pass 8 instrument response functions\footnote{\href{https://fermi.gsfc.nasa.gov/ssc/data/analysis/documentation/Cicerone/Cicerone_Data_Exploration/Data_preparation.html}{fermi.gsfc.nasa.gov/ssc/data/analysis/documentation/Cicerone/\\Cicerone\_Data\_Exploration/Data\_preparation}}. Good time intervals are selected using \texttt{gtmktime} using the standard filter \texttt{(LAT\_CONFIG==1 \&\& DATA\_QUAL$>$0)}. Additionally, only time intervals during which the Sun is at least 5\degr\ away from the region of interest are kept, so as to avoid contamination by potential solar flares. The Moon being a bright $\gamma$-ray emitter as well \citep{2016PhRvD..93h2001A}, data when the Moon is closer than 5\degr\ from the region of interest could also be filtered out as recommended by \citet{2013arXiv1302.5141C}, if the input spacecraft file has been previously processed with the \texttt{moonpos} script\footnote{\href{https://fermi.gsfc.nasa.gov/ssc/data/analysis/user/moonpos-1.1.tgz}{https://fermi.gsfc.nasa.gov/ssc/data/analysis/user/moonpos-1.1.tgz}}. This last procedure is left to the discretion of the user. The evolution of the count rate is then computed using the \texttt{LC} method of \texttt{gtbin}. To correct for the time-dependent exposure on a source with \texttt{gtexposure} and obtain a flux, a model of the source of interest has to be provided. To this end, the user-contributed script \texttt{make3FGLxml} is used, thus accounting for all sources included in the 3FGL catalogue \citep{2015ApJS..218...23A} near the source of interest. If the source of interest does not belong to the 3FGL catalogue, a photon index of 2.5 is assumed. During the process of light curve computation with \texttt{FLaapLUC}, the photon index of the source of interest is thus set constant, with a value either corresponding to the 3FGL catalogue or fixed to 2.5. No potential spectral temporal variation is considered.

\texttt{FLaapLUC} can then be used to generate triggers in two different ways:
\begin{enumerate}
  \item One can manually define a fixed flux threshold, on a source by source basis;
  \item Alternatively, if a long-term light curve using all the available \textit{Fermi}-LAT data has been pre-computed, \texttt{FLaapLUC} can dynamically assess the flux threshold above which a source activity will generate a trigger (see below).
\end{enumerate}

In the latter case, such a pre-computed long-term light curve can typically be generated using the following command:
\begin{verbatim}
  flaapluc --merge-long-term
      --config-file=config/<config_file.cfg>
      <source name>
\end{verbatim}
where {\verb?config_file.cfg?} is a configuration file where several options can be set. An example configuration file is provided in {\verb?FLaapLUC/config?} on GitHub.

\begin{figure}
  \centering
  \includegraphics[width=\columnwidth]{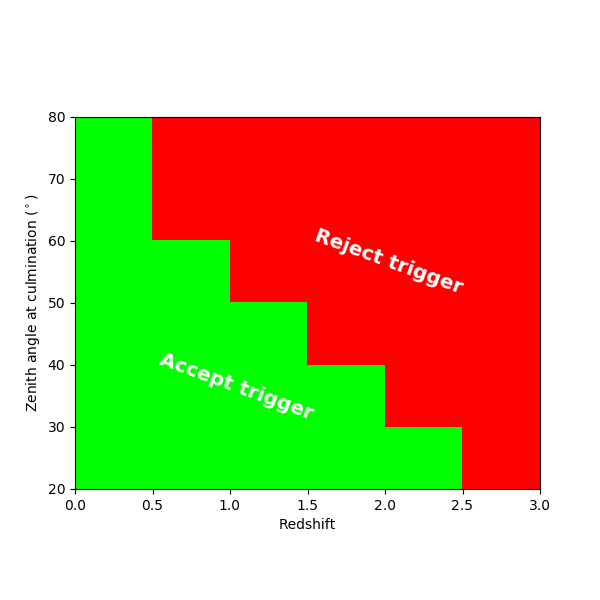}
  \caption{Two-dimensional criterion applied on the redshift and the zenith angle at culmination of a source to accept or veto a latent alert from \texttt{FLaapLUC}. These specific cut values are used in H.E.S.S. for the \textit{Fermi}-LAT data analysis of extragalactic sources.}
  \label{fig:maskZAz}
\end{figure}

The main difficulty in the procedure resides in the definition of a flare, while the event is actually ongoing. Specifically, how to dynamically compute such a flux threshold with respect to the long-term average flux level. Even once the entire data set is available, the definition of what a flare is, is subject to debate. For instance, \citet{2013MNRAS.430.1324N} proposes that a flare consists in finding the peak in a light curve and to consider the contemporaneous temporal window when the flux is at least half of the peak flux. Here, one can not use such a definition, which requires having the full observation data set at hand. Indeed, the aim of \texttt{FLaapLUC} is to alert for an \emph{ongoing} event, without knowing whether the last flux measurement still corresponds to a trend of rising flux, or whether the flare is already on its decay. Instead, the following approach is proposed. For a given source, a weekly-binned long-term light curve is pre-computed, thus currently using more than 9 years of \textit{Fermi}-LAT data. To assess whether the source is experiencing an active state, another light curve is computed every day, using bins of $N_1$ days of duration, and the last flux bin measurement is compared to the long-term flux mean. If the last flux bin is significantly higher than the average flux, \texttt{FLaapLUC} identifies the current flux state as \emph{active}. Following this, another, finer, light curve is generated with bins of $N_2$ days (with $N_2 < N_1$). If the new more finely binned flux is significantly above the long-term flux mean, \texttt{FLaapLUC} issues an alert.

More quantitatively, and accounting for flux errors, the averaged flux ($\overline{F_{LT}}$) on long-term data is computed.
Let's denote $F_{N_{1,2}}$ the last $N_{1,2}$-days binned flux value, and $\delta F_{N_{1,2}}$ its error.
A two-level criterion on the flux is set, based on the $N_1$-days binned and $N_2$-day binned light curves. The trigger threshold on the flux is such that:

\begin{align}
  F_{N_1} - \delta F_{N_1} &> \overline{F_{LT}} + \alpha_{N_1} \; \text{RMS}(F_{LT}) \notag \\
  F_{N_2} - \delta F_{N_2} &> \overline{F_{LT}} + \alpha_{N_2} \; \text{RMS}(F_{LT})
  \label{eq:trig_flux}
\end{align}

To speed the processing of potentially many sources every day, the $N_2$-day binned light curve is only computed if the first criterion on $F_{N_1}$ is fulfilled. The trigger threshold, and thus probability (see Section\ref{sec:falsealarm}), then depends on the settings on $\alpha_{N_1}$ and $\alpha_{N_2}$, which are chosen by the user so as to tune the alert rate for a particular source class. The chosen value of $N_2$ thus limits the minimum time scale of a flare \texttt{FLaapLUC} can probe in \textit{Fermi}-LAT data. Prompt alerts are further limited by the latency time required to downlink the data and reconstruct them. This will be further discussed in Section~\ref{sec:latency}. In any case, it is unfeasible to react on the fly to rapid events such as the giant outburst detected from 3C\,279 in June 2015 \citep{2016ApJ...824L..20A}, which exhibited doubling times of less than 5 minutes, or the equivalent at high energies of the very high energy flares of PKS\,2155$-$304 seen in 2006 \citep{2007ApJ...664L..71A} or Mrk\,501 as observed in 2005 \citep{2007ApJ...669..862A} which varied on similar time scales.

Moreover, if one wants to monitor the high energy sky in order to trigger ToO observations at a particular site, it is useful to additionally filter alerts on the visibility of the sources in the next hours/days. \texttt{FLaapLUC} can thus perform such a filtering, depending on the visibility of a source at a given site and observation time, using the \texttt{pyephem} package\footnote{\href{http://rhodesmill.org/pyephem}{rhodesmill.org/pyephem}}. As an example, the common set of trigger criteria used within the H.E.S.S. extragalactic working group is the following:

\begin{itemize}
\item the source should have its last flux measurement fulfilling the criteria described in Eq.~\ref{eq:trig_flux}, with $N_1=3$ days, $N_2=1$ day, $\alpha_{N_1}=2$ and $\alpha_{N_2}=3$;
\item the source should be visible the next night at the H.E.S.S. site (Lon. 23\degr16\arcmin18\arcsec S, Lat. 16\degr30\arcmin00\arcsec E), and its zenith angle at culmination should be less than a certain value which depends on the redshift of the source, due to the absorption by the extragalactic background light \citep{2001ARA+A..39..249H} of the observed source spectrum at very high energies.
\end{itemize}

\begin{figure}
  \centering
  \includegraphics[width=\columnwidth]{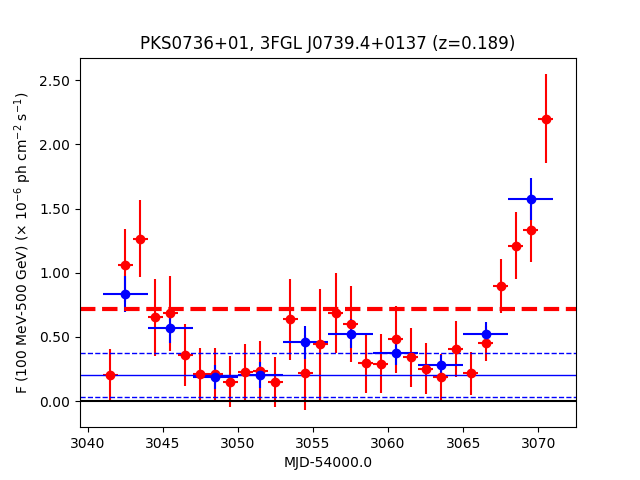}
  \caption{\texttt{FLaapLUC} light curve output on PKS\,0736$+$01 issued on Feb. 18, 2015. The blue points show the 3-day binned light curve, while the red points show the daily-binned one. The horizontal blue line represents the long-term flux average of the source, and the horizontal dotted blue lines are the flux levels plus or minus $2 \cdot \text{RMS}(F_{LT})$ away from this average ($\alpha_{N_1}=2$). The horizontal bold red line shows the flux threshold for $\alpha_{N_2}=3$ above which \texttt{FLaapLUC} issues an alert on this source.}
  \label{fig:flaaplucexample}
\end{figure}

\begin{figure}
  \centering
  \includegraphics[width=\columnwidth]{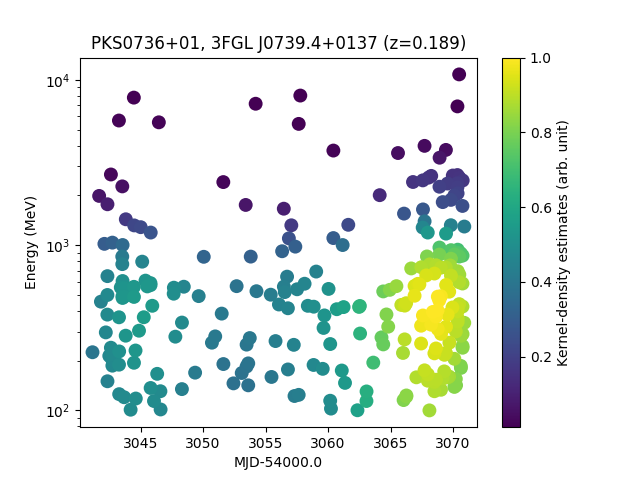}
  \caption{Energy versus arrival time plot on PKS\,0736$+$01 of each individual \textit{Fermi}-LAT event as of Feb. 18, 2015. The colour code depicts a simple Gaussian kernel-density estimate for visualisation purposes. Qualitatively, a vertical clustering of yellow points would denote a flare.}
  \label{fig:energytime}
\end{figure}

The reasoning behind the last criterion is the following. The very high energy $\gamma$-ray photons experience absorption on their propagation path due to the extragalactic background light \citep{2001ARA+A..39..249H}. This absorption depends on the photon energy, and is more severe at the highest energies which imaging atmospheric \v{C}erenkov telescopes (IACT) are sensitive to. Since the energy threshold of IACT also increases with the observation zenith angle, for a similar flux, further away sources should be observed at smaller zenith angles (i.e. higher elevation) than closer ones to reach the same detection probability. This last cut is modular and programmable. It can be implemented as a simple scalar value on the maximal acceptable zenith angle and/or redshift, or can be mapped as a two-dimensional criterion, as depicted in Fig.~\ref{fig:maskZAz}. For sources whose redshift is unknown, a value of $z=0$ is used (here usually the most permissive zenith angle at culmination is considered).

\texttt{FLaapLUC} takes photon files from \textit{Fermi}-LAT data as input. The pipeline can be run using e.g. an all-sky photon file encompassing all the $\gamma$-ray-like events recorded with \textit{Fermi} LAT for the whole mission. This is necessary in order to pre-compute a long-term light curve for multiple sources at once. Alternatively, for a daily running, one can use an all-sky file from a subset of the last data acquired with \textit{Fermi}-LAT to speed up the computation and limit the input/output usage in case many sources are to be processed. A roll back time of typically one month is used by the H.E.S.S. collaboration. Such input files can easily be generated, as well as an automatic retrieval of spacecraft and photon files, using for instance \texttt{enrico} \citep{2013arXiv1307.4534S}. \texttt{FLaapLUC} is actually using \texttt{enrico} to generate those input files on the fly in case the user does not provide them. The daily running instance of \texttt{FLaapLUC} is typically run using the following command:

\begin{figure*}
  \centering
  \includegraphics[width=\textwidth]{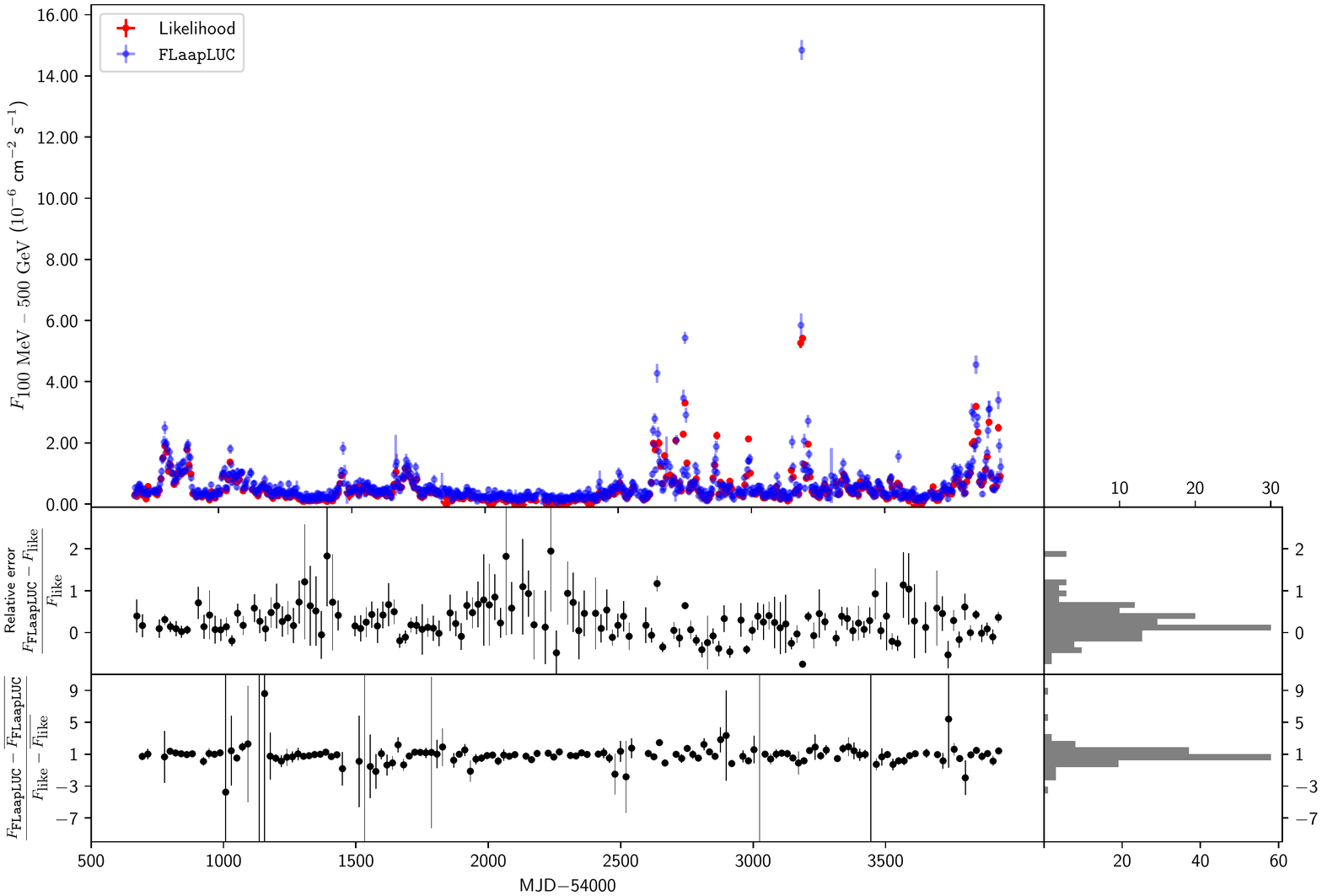}
  \caption{\textit{Top}: Example of a long-term light curve computed with \texttt{FLaapLUC} for 3C\,279 in blue, and the same computed with the binned likelihood approach in red. For instance, the June 2015 flare is well visible. \textit{Middle left}: Relative errors of the aperture photometry analysis with respect to the likelihood results. \textit{Middle right}: Distribution of the relative errors. \textit{Bottom left}: Ratio between the aperture photometry and the likelihood results, once debiased from their respective average. \textit{Bottom right}: Corresponding distribution.}
  \label{fig:3C279LT}
\end{figure*}

\begin{figure*}
  \centering
  \includegraphics[width=\textwidth]{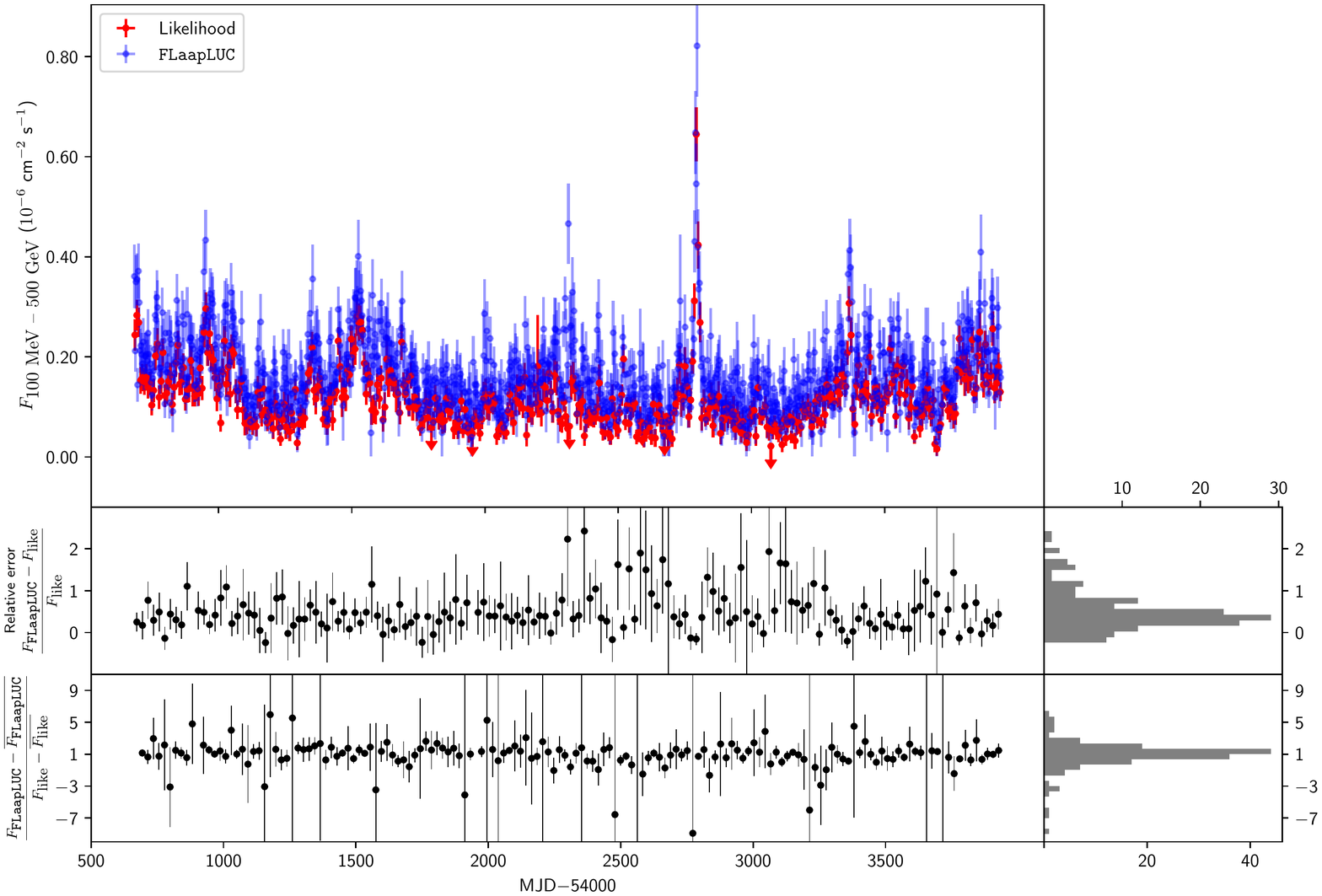}
  \caption{Same as Fig.~\ref{fig:3C279LT} for PKS\,2155$-$304.}
  \label{fig:2155LT}
\end{figure*}

\begin{verbatim}
  flaapluc-allsources --daily
      --custom-threshold
      --with-history
      --config-file=config/<config_file.cfg>
      <source list file>
\end{verbatim}
where \texttt{flaapluc-allsources} is just a wrapper of the script \texttt{flaapluc} looping through all the sources listed in the input file.

Within the H.E.S.S. collaboration, whenever \texttt{FLaapLUC} issues an alert, with the last bit of data on a monitored source fulfilling the double-pass threshold from Eq.~\ref{eq:trig_flux}, as well as constraints on visibility, redshift and zenith angle at culmination, a more detailed likelihood analysis is automatically performed on the last $N_1$ days of available data. This procedure leads to a computational economy with respect to a scheme where all monitored sources would have been analysed with the full likelihood approach.

\texttt{FLaapLUC} has been used in H.E.S.S. since 2012 successfully, having given rise to quick reaction follow-ups, such as in February 2015 on PKS\,0736$+$01 (see Fig.~\ref{fig:flaaplucexample}) which resulted in the detection of this source during this event with H.E.S.S. at very high energies \citep{2016arXiv161005523C}. \texttt{FLaapLUC} issued an alert even before public information was available on this flare. Figure~\ref{fig:energytime} shows the energy versus arrival time for each event from the alert on PKS\,0736$+$01. The colours depict the density of events in the data with a Gaussian kernel-density estimate using \texttt{scipy.stats.gaussian\_kde}. This also allows an assessment of the energy of the highest energy photon received during a flaring event. Such information can be useful for deciding whether or not ToO observations should be triggered at higher energies, with e.g. the H.E.S.S. experiment.

Apart from AGN, \texttt{FLaapLUC} is also used internally in H.E.S.S. to produce alerts on a predefined list of $\gamma$-ray binaries or binary candidates, in this case using different criteria on the flux thresholds and observability, with $N_1=2$ days, $N_2=1$ day, $\alpha_{N_1}=2$, $\alpha_{N_2}=3$ and a fixed maximum allowed zenith angle at culmination of 60\degr.

As a third application, a systematic survey of the Galactic plane is performed daily at high energies with \texttt{FLaapLUC}, with a scan of 540 regions of 1\degr\ of radius, in the Galactic latitude band $|b| < 3\degr$. Again, even though the Galactic plane is largely dominated by the Galactic diffuse emission which thus hampers any absolute flux determination with the aperture photometry method, any significant relative flux variation could be detected with this tool. For this application, the trigger criteria are $N_1=2$ days, $N_2=1$ day, $\alpha_{N_1}=3$, $\alpha_{N_2}=5$ and a fixed maximum allowed zenith angle at culmination of 60\degr.

\section{Performance}
\label{sec:perf}

The performance and limitations of \texttt{FLaapLUC} are hereafter developed. Table~\ref{tab:summary} gives a summary of the main points discussed in this paper, as well as the operational settings used by H.E.S.S. as mentioned above.

\begin{table*}
  \caption{Summary of the operational settings, performances and limitations of \texttt{FLaapLUC} for the applications in use in H.E.S.S.}
  \label{tab:summary}
  \centering
  \begin{tabular}{lccc}
    \hline\noalign{\smallskip}
                              & Extragalactic sources & Binary candidates & Galactic plane survey\\
    \noalign{\smallskip}\hline\noalign{\smallskip}
    $N_1$                     & 3 days                & 2 days            & 2 days\\
    $\alpha_{N_1}$            & 2                     & 2                 & 3\\
    $N_2$                     & 1 day                 & 1 day             & 1 day\\
    $\alpha_{N_2}$            & 3                     & 3                 & 5\\
    False alarm probability   & $<0.3\%$              & $<0.3\%$          & $<0.05\%$\\
    Minimum time scale probed & $\sim$ 1 day          & $\sim$ 1 day      & $\sim$ 1 day\\
    \noalign{\smallskip}\hline
  \end{tabular}
\end{table*}

\subsection{Comparison with the likelihood method}

The aperture photometry method provides a fast way to obtain relative results, but is obviously not the best choice when it comes to reliable, absolute flux measurements, because of the basic procedure consisting in attributing all photons from an analysed region to a given source of interest. In this section, the light curve obtained using \texttt{FLaapLUC} is compared to the one computed using the binned likelihood scheme, for the same object. Data obtained from August 4, 2008 to July 4, 2017 are analysed, for two sources, 3C\,279 (a flat spectrum radio quasar, FSRQ) and PKS\,2155$-$304 (a high-frequency-peaked BL\,Lac object, HBL), as an illustration.

For the likelihood analyses, events in a region of interest of 10\degr\ radius were selected. The \texttt{PASS 8} instrument response functions (event class 128 and event type 3) corresponding to the \texttt{P8R2\_SOURCE\_V6} response were used together with a zenith angle cut of 90\degr. The model of the region of interest was built based on the 3FGL catalogue \citep{2015ApJS..218...23A}. The Galactic diffuse emission has been modeled using the file \texttt{gll\_iem\_v06.fits} \citep{2016ApJS..223...26A} and the isotropic background using \texttt{iso\_P8R2\_SOURCE\_V6\_v06.txt}. The fit is performed iteratively: in a first step, sources from the 3FGL catalogue within 15\degr\ around the source of interest are included, with parameters fixed for those more than 10\degr\ away to account for the large point spread function at low energies. In a second step, parameters of sources contributing to less than a test statistic \citep[TS,][]{1996ApJ...461..396M} of 9 and to less than 5\% of the total number of counts in the region of interest are frozen. In a third step, the only free parameters are those of sources less than 3\degr\ away from our source of interest (if not frozen in the previous step), the source of interest itself, and the normalisations of the Galactic and isotropic diffuse emissions. In the different steps, the spectral parameters (photon and curvature indices, since both PKS\,2155$-$304 and 3C\,279 are described with log-parabolic spectra in the 3FGL catalogue) are fixed to the catalogue values. This is to ensure a proper comparison with \texttt{FLaapLUC} results, since the latter does not account for potential spectral evolution as a function of time.

The results of the likelihood analyses of 3C\,279 and PKS\,2155$-$304 are shown in Figs.~\ref{fig:3C279LT} and \ref{fig:2155LT} respectively. Weekly-binned light curves are shown in the top panel for both \texttt{FLaapLUC} results and the likelihood analysis. The middle left panel represents the relative error between the two analysis methods and the middle right panel displays the distribution of this error. \texttt{FLaapLUC} systematically overshoots the resulting flux compared with a proper likelihood analysis, which is inherent to the original assumption of a data set free of any background (see middle right panel in Fig.~\ref{fig:3C279LT}). This is especially the case for low fluxes (e.g. for 3C\,279, see middle left panel in Fig.~\ref{fig:3C279LT}, e.g. around MJD~55300 or MJD~56000--56300) where the contribution from the diffuse emission components (Galactic and extragalactic) is not negligible at all. The error distributions show that the aperture photometry overestimates the fluxes by $\sim 30$--$50\%$ on average, and up to a factor $\sim 2$ in case of low activity. However, it can be seen from Figs.~\ref{fig:3C279LT} and \ref{fig:2155LT} that the global trends of the light curves are well reproduced in the aperture photometry results compared to the likelihood ones. This is further strengthened in the bottom panels, which show a comparison of the \texttt{FLaapLUC} and the likelihood results once debiased from their respective average.

\subsection{False alarm rate}
\label{sec:falsealarm}

\begin{figure}
  \centering
  \includegraphics[width=\columnwidth]{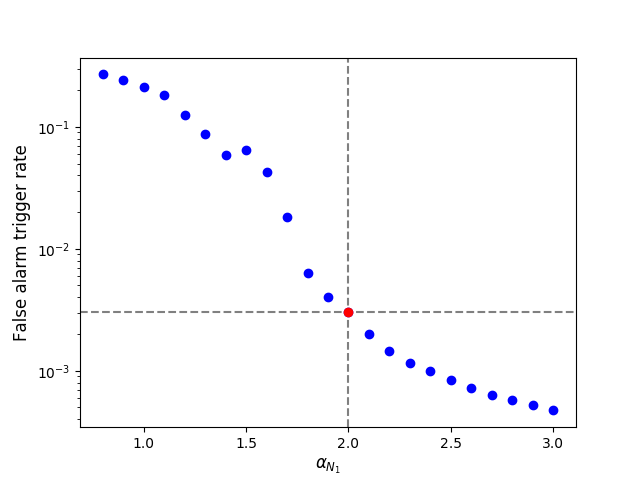}
  \caption{False alarm trigger probability on the first iterative light curve generation step as a function of $\alpha_{N_1}$. The setting used for extragalactic sources in H.E.S.S. is shown with the dotted gray lines and red point.}
  \label{fig:falsealarm}
\end{figure}

The false alarm probability of the \texttt{FLaapLUC} pipeline is determined by simulating light curves of AGN following \citet{2013MNRAS.433..907E}. To do so, a scan is performed through the $\alpha_{N_1}$ parameter using $N_1=3$ days to assess when \texttt{FLaapLUC} would generate a finer binned light curve of $N_2$ days. First, for each scanned value of $\alpha_{N_1}$, sources from the monitored list which have never triggered \texttt{FLaapLUC} so far in the first iterative step on the flux criterion were identified. For each of those sources, 1000 mock light curves were simulated preserving the underlying probability density function and power spectral density from real data using the method described in \citet{2013MNRAS.433..907E} and adapted to \texttt{Python} by \citet{2015arXiv150306676C}. \texttt{FLaapLUC} was run using the simulated light curves as inputs, and the false alarm probability on $\alpha_{N_1}$ was taken from the rate at which $N_2$-day binned light curve are generated. Figure~\ref{fig:falsealarm} presents the false alarm probability when varying the threshold parameter $\alpha_{N_1}$. The operation point used in the H.E.S.S. extragalactic working group is shown in red, corresponding to wrongly generating the finer light curve in 0.3\% of the cases. The false alarm probability of the whole double-pass procedure is not evaluated, being too resource consuming to be properly computed. However, it seems safe to state that with such running settings on $\alpha_{N_1}=2$ and $\alpha_{N_2}=3$, this false alarm rate is well below $\sim 0.1\%$ for the AGN monitored within the H.E.S.S. collaboration.

\subsection{Computing time and latency}
\label{sec:latency}

Processing a single source with \texttt{FLaapLUC} takes slightly more than 5 minutes, when the input data set is limited to the last 30 days of available data. In H.E.S.S., the \texttt{FLaapLUC} production instance monitors about 900 sources every day, including about 300 AGN, 540 regions in the Galactic plane, and about 60 binaries and other Galactic sources. The whole processing typically takes 2\,hr (wall clock time) when 60 concurrent jobs are run at the IN2P3 computing cluster\footnote{\href{https://cc.in2p3.fr/en}{cc.in2p3.fr/en}} (CC-IN2P3).

As stated above, to organise follow-up observations on a flare, a prompt reaction is essential. Summing up the time needed to transfer the data from the \textit{Fermi} spacecraft to the ground, to digest them and then retrieve photon and spacecraft pointing files from the NASA servers, to generate an all-sky file which is used as input for \texttt{FLaapLUC}, and to analyse all the monitored sources, the total latency time of the process, i.e. the delay between the last bit of data and the generation of an alert with \texttt{FLaapLUC}, is about 8\,hr. However, since this daily processing is usually run between 3:30 UTC and 6:00 UTC in H.E.S.S., some time is left to assess whether ToO observations should be triggered with H.E.S.S. for the next night, and with other multiwavelength facilities.

\section{Conclusions and prospects}
\label{sec:ccl}

\texttt{FLaapLUC}, a tool designed to provide alerts on the fly on transient high energy sources using \textit{Fermi}-LAT data, was presented. This pipeline can provide quick results to allow the prompt organisation of follow-up, multiwavelength observations. The method is based on aperture photometry, which is not well suited to provide absolute flux measurements of \textit{Fermi}-LAT data, but can be used to assess relative time variations from high energy emitting objects.

It has been shown that \texttt{FLaapLUC} is quick and efficient, and thus useful for providing alerts on flaring events from \textit{Fermi}-LAT data. This can help in the organisation of follow-up ToO observations of transient $\gamma$-ray sources, for example with IACT such as VERITAS, MAGIC and H.E.S.S.

\texttt{FLaapLUC} results are compared to full likelihood analyses, which show good agreement on relative flux variations. An evaluation of the associated false alarm probability reveals that this tool is robust and efficient to detect transient events. A limitation comes from the latency of the overall data processing, of about 8\,hr, preventing the possibility of generating useful prompt alerts on events occurring on shorter time scales.

As long as the background is approximately constant, the aperture photometry method can be used to quickly detect active states from sources in data acquired by any instrument producing event lists. For instance, it is conceivable to adapt such a system for online triggering alerts for the future CTA observatory \citep{2013APh....43....3C}, if events could be reconstructed fast enough \citep[see e.g.][]{2014SPIE.9145E..2XB}.

\texttt{FLaapLUC} \citep{2017ascl.soft09011L} has been made publicly available on GitHub at \href{https://github.com/jlenain/flaapluc}{github.com/jlenain/flaapluc}, and contributions from the community are warmly welcome.

\section{Acknowledgements}
I am very thankful to my colleagues within the H.E.S.S. collaboration for very fruitful discussions which led to the implementation of \texttt{FLaapLUC} and further improvements, and especially to Stefan Wagner, Michael Punch, Heike Prokoph, Matteo Cerruti, Bruno Kh\'elifi, Pol Bordas, V{\'i}ctor Zabalza and Julien Bolmont. I am grateful to Agnieszka Jacholkowska, Matteo Cerruti, Julien Bolmont and Andrew Taylor for their careful reading of this manuscript. I would like to thank the anonymous referee for constructive inputs.

This research made use of \texttt{Enrico}, a community-developed Python package to simplify \textit{Fermi}-LAT analysis \citep{2013arXiv1307.4534S}. This research has made use of NASA's Astrophysics Data System. This research has made use of the SIMBAD database, operated at CDS, Strasbourg, France \citep{2000A+AS..143....9W}. I gratefully acknowledge CC-IN2P3 (\href{https://cc.in2p3.fr}{cc.in2p3.fr}) for providing a significant amount of the computing resources and services needed for this work.

This work is dedicated to the memory of my missed friend, Jean-Claude Rouffignat, who introduced me to astronomy when I was a child. You are not forgotten.

\bibliographystyle{elsarticle-names}
\bibliography{flaapluc.bbl}

\end{document}